# HUMAN IMMUNODEFICIENCY VIRUS (HIV) CASES IN THE PHILIPPINES: ANALYSIS AND FORECASTING

**Analaine May A. Tatoy and Roel F. Ceballos**

Department of Mathematics and Statistics

College of Arts and Sciences

University of Southeastern Philippines

Obrero, Davao City, Philippines

## Abstract

Reports from the Health Department in the Philippines show that cases of Human Immunodeficiency Virus (HIV) are increasing despite management and control efforts by the government. Worldwide, the Philippines has one of the fastest growing number of HIV cases. The aim of the study is to analyze HIV cases by determining the best model in forecasting its future number of cases. The data set was retrieved from National HIV/AIDS and STI Surveillance and Strategic Information Unit (NHSSS) of the Department of Health containing 132 observations. This data set was divided into two parts, one for model building and another for forecast evaluation. The original series has an increasing trend and is nonstationary with indication of non-constant variance. Box-Cox transformation and ordinary differencing were performed on the series. The differenced series is stationary and tentative models were identified through ACF and PACF plots. SARIMA has the smallest chosen AIC value. The chosen model undergoes the diagnostic checking. The residuals of the model behave like a white noise while the forecast errors behave like a Gaussian white noise. Considering all diagnostics, the model may be used for forecasting the monthly cases of HIV in the Philippines. Forecasted values show that HIV cases will maintain their current trend.

## 1. Introduction

Human Immunodeficiency Virus (HIV), the virus that causes AIDS (Acquired Immunodeficiency Syndrome), has become one of the world's most serious health and wellness challenges. According to Joint United Nations Programme on HIV/AIDS (UNIAIDS), 77.3 million people have become infected with HIV since the start of the epidemic in year 1981. In 2017, approximately 36.9 million people were currently living with HIV. HIV is a virus spread through certain body fluids that attack the body's immune system, specifically the CD4 cells (T cells) which damages the immune system and makes it harder and harder for the body to fight infections and some other disease (Center for Disease Control   and Prevention [1]).

Over the past decades, major global efforts have been mounted to address the epidemic, and significant progress has been made. In 2017, 21.7 million people have access to antiretroviral therapy (UNIAIDS [11]). Although, HIV testing capacity has increased over time, enabling more people to learn their HIV status, HIV remains a leading cause of death worldwide. The Sub-Saharan Africa, with more than two-thirds of all people living with HIV globally, is the hardest hit region in the world, followed by Asia and the Pacific. The Caribbean, as well as Eastern Europe and Central Asia are also heavily affected (UNIAIDS [11]).

The Philippines has registered the fastest-growing HIV/AIDS epidemic in the Asia-Pacific region in the past six years with a 140-percent increase in the number of new infections (Manila Rueters [4]). The first Human Immunodeficiency Virus (HIV) case in the Philippines was reported in 1984. Around 2007, the DOH noted a rise in epidemic as new infections started showing a steady spike and shifted from sex workers to men who have sex with men (MSM) and people who inject drugs (Santos [9]). The number of HIV/AIDS in the country continues to increase as the Department of Health (DOH) has documented a total of 11,103 cases in 2017 (Philstar [7]). In April 2017, 629 new persons, most of whom are millennial's, were diagnosed with the HIV and more than 80 percent or 513 of those diagnosed with the virus belong to the 15 to 34 age group (CNN, Philippines [2]).

Due to the increasing number of reported HIV cases in the Philippines, there is a need to analyze the trend and utilize existing methods to forecast future values of HIV cases in order to have a basis for the creation of necessary intervention programs and appropriate health policies. Time series methods aside from being utilized in economic data are also used in the analysis and modeling of health time series data such as Malaria incidence. The aim of this study is to find an appropriate time series model for HIV cases in the Philippines using the Box-Jenkins approach. The Box-Jenkins approach has been widely used in different fields such as economics (Manayaga and Ceballos [2]) and health (Perez and Ceballos [6]).

## 2. Methodology

The Box-Jenkins forecasting method consists of a three-iterative procedure as follows: model identification, parameter estimation and diagnostic checking. In addition, a step called *forecast evaluation* is also suggested. R is a programming language and environment which provides a variety of statistical operations and graphics. It was first created in New Zealand by Ross Ihaka and Robert Gentleman of the University of Auckland, and the software is currently developed by R development Core Team. The researcher used the R software in producing plots and running the following R packages such as; 'tseries' for testing stationarity; 'astsa' for getting the numerical values of ACF and PACF in model identification; 'forecast' for model estimation and forecasting; 'lm test' for checking of parameter significance; 'stats' for generating the L-Jung Box test and some built-in R packages as well that will help run the analysis.

## 3. Results and Discussion

Figure 1 displays the time series plot of the monthly number of Human Immunodeficiency Virus cases in the Philippines. The time series data is composed of 108 monthly observations from January 2009 to December 2016. Generally, the time series plot exhibits an increasing trend. There is also an indication that the variance is not constant. That is, the variability in some periods appears to be larger than the other periods. Based on the plot shown in Figure 1, there is an existing problem in the variance of the original data. To deal with this, Box-Cox transformation is performed to stabilize the variance. Figure 2 displays the time series plot of the transformed data using the optimal lambda of 0.490. An increasing trend is still present in the series which implies that the series is nonstationary even after transformation.

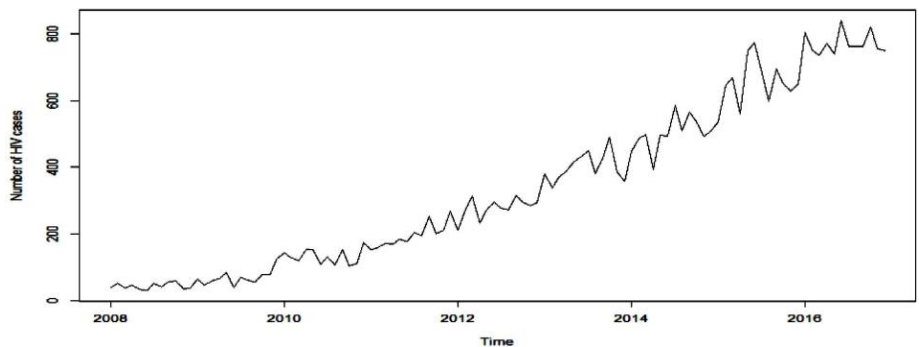

**Figure 1.** Time series plot of HIV cases.

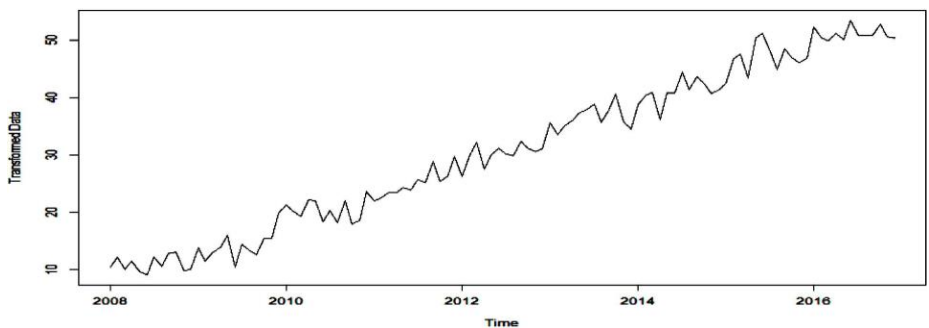

**Figure 2.** Time series plot of the transformed series.

To eliminate the nonstationary of the transformed series, difference is necessary. Hence, first order differencing was applied to the transformed series and the result is shown in Figure 3. It can be observed that most of the data points fluctuate randomly around a constant mean which indicates that the series is now stationary. To formally test the stationarity of the difference of the transformed series, the Augmented Dickey-Fuller (ADF) test is applied. Table 1 shows that the ADF value is –5.5074 with a corresponding $p$-value of less than 0.01. Since its $p$-value is less than 0.05, the null hypothesis is rejected. This implies that the first difference of the transformed series is stationary. Thus, we can now proceed with model identification using the ACF and PACF plots.

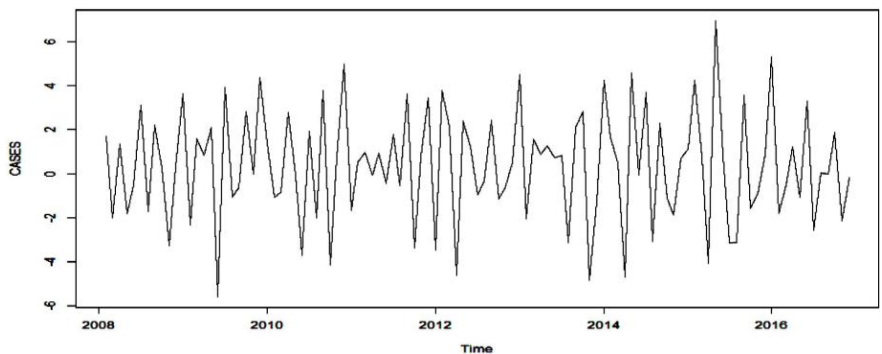

**Figure 3.** Time series plot of HIV cases.

**Table 1.** ADF test for stationarity of the first difference

| Test statistic | $p$-value |
|---|---|
| –5.5074 | < 0.01 |

To determine the set of tentative models, the ACF and PACF plots of the stationary series are presented in Figure 4. In terms of seasonal lags (lags at period 12, 24, 36, and so on), the ACF cuts off after lag 12 and PACF tails off. This results in a model with no seasonal autoregressive (AR) term and one seasonal moving average (MA) term. Another possibility might be a PACF cut off after lag 12 and an ACF that tails off. This results in a model with no seasonal MA term and one seasonal AR term. In terms of nonseasonal lags, the ACF and PACF exhibit a mixture of exponential decay and damped sinusoid. This is an indication that the model contains both AR and MA terms. Another possibility might be an ACF that cuts off after lag 1 and a PACF tails off. This results in a model with one MA term and no AR terms. Considering all combinations of these interpretations, Table 2 presents the list of tentative models for the monthly HIV cases in the Philippines together with their respective AIC values. We may consider $SARIMA(2, 1, 1) \times (1, 0, 0)_{12}$, $SARIMA(2, 1, 2) \times (1, 0, 0)_{12}$ and $SARIMA(0, 1, 1) \times (1, 0, 0)_{12}$ as best models since these have least AIC values among the 10 tentative models. However, the $SARIMA(0, 1, 1) \times (1, 0, 0)_{12}$ is the selected model since it has the least AIC value of 469.2248 and it yielded significant result among all its coefficients.

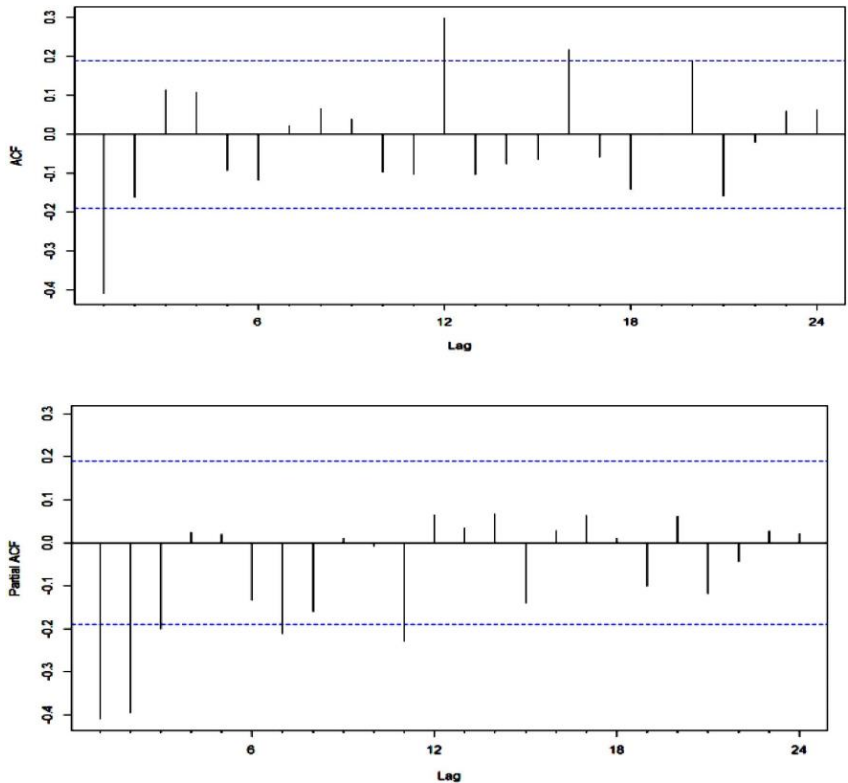

**Figure 4.** ACF and PACF plots of the stationary series.

**Table 2.** The overall utility estimates of extrinsic attributes

| Models | AIC |
|--------|-----|
| SARIMA$(0, 1, 1) \times (1, 0, 0)_{12}$ | 469.2248 |
| SARIMA$(0, 1, 1) \times (0, 0, 1)_{12}$ | 470.7467 |
| SARIMA$(1, 1, 1) \times (1, 0, 0)_{12}$ | 471.0778 |
| SARIMA$(2, 1, 2) \times (0, 0, 1)_{12}$ | 471.9348 |
| SARIMA$(1, 1, 2) \times (1, 0, 0)_{12}$ | 472.3695 |
| SARIMA$(1, 1, 2) \times (0, 0, 1)_{12}$ | 474.2011 |
| SARIMA$(2, 1, 1) \times (1, 0, 0)_{12}$ | 469.199 |
| SARIMA$(2, 1, 1) \times (0, 0, 1)_{12}$ | 471.251 |
| SARIMA$(1, 1, 1) \times (0, 0, 1)_{12}$ | 472.6633 |
| SARIMA$(2, 1, 2) \times (1, 0, 0)_{12}$ | 470.5647 |

The estimates of the model's coefficients for SARIMA$(0, 1, 1) \times (1, 0, 0)_{12}$ are presented in Table 3 along with their standard error, $z$-values and $p$-values. It can be observed that MA(1) has a value of –0.557835 and SAR(1) coefficient has a value of 0.408311. These two coefficients have $p$- values less than 0.05. Hence, both coefficients are significantly different from zero.

**Table 3.** Model estimates of SARIMA$(0, 1, 1) \times (1, 0, 0)_{12}$

| Model term | Estimates | Standard error | $z$-value | $p$-value |
|------------|-----------|----------------|-----------|-----------|
| MA(1) | –0.557835 | 0.074869 | –7.4508 | < 0.01 |
| AR(1) | 0.408311 | 0.091364 | 4.4691 | < 0.01 |

To check the adequacy of the model, residual analysis is performed. The residuals of the selected model were obtained and plots are presented in Figure 5. The plot of residuals against the fitted values exhibits no strong patterns (Figure 5(a)).

Also, residuals are structure less when plotted against time (Figure 5(b)). This is an indication that residuals have a constant variance and are independent. Furthermore, the normal probability plot of residuals shows that majority of the residuals lie along the theoretical line of normality (Figure 5(c)). Hence, the residuals are assumed to be normally distributed. These plots are used to assess how well the chosen model fits the data.

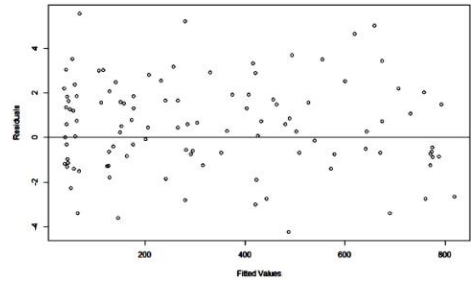

**5(a)**

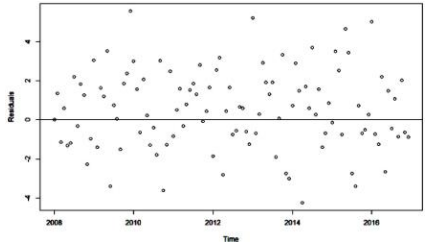

**5(b)**

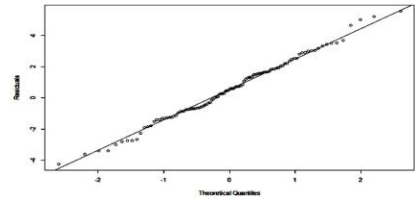

**5(c)**

**Figure 5**

To further diagnose the chosen model, ACF and PACF plots are obtained and presented in Figure 6. There is a significant spike at lag 16 for ACF plot. However, majority of the ACF and PACF values are within acceptable limits and are not significant. This means that the residuals behave like a white noise series.

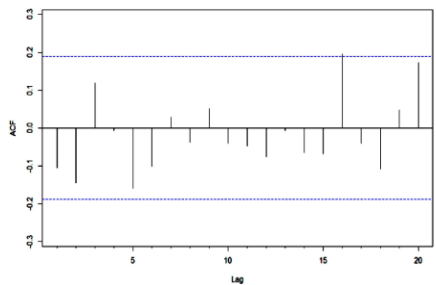

**6(a).** PACF residuals.

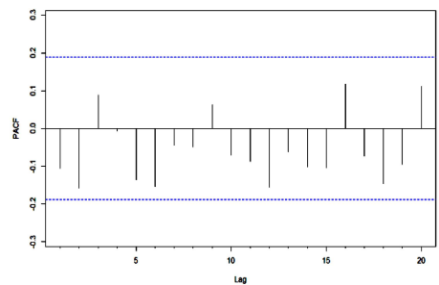

**6(b).** ACF residuals.

**Figure 6**

To formally test whether the residuals of the chosen model are uncorrelated, Ljung-Box test was applied. Table 4 shows the Ljung-Box test statistic of 23.172 with a *p*-value of 0.1841. Since the *p*-value is greater than 0.05 level of significance, there is not enough evidence to say that serial autocorrelation exists among the residuals. Hence, the model is appropriate for the series. Moreover, based on the result of Ljung-Box test and the behavior of the residuals in the previous plot, the residuals are possibly generated by a white noise process.

**Table 4.** Ljung-Box test for residuals

| Test statistic | df | *p*-value |
|---|---|---|
| 23.172 | 18 | 0.1814 |

The one-step ahead forecast for the month of January 2017 up to the month of December 2018 (24 months) is performed and these forecasted values are compared with the actual values to obtain the forecast error. The ACF and PACF plots of the forecast errors show that there is a significant spike at lag 1 for both plots. However, majority of the values are within acceptable limits which suggest that the forecast errors are white noise.

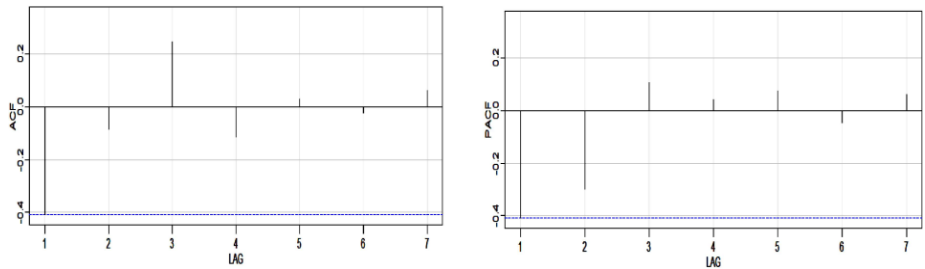

**7(a).** ACF forecast errors.          **7(b).** PACF forecast errors.

**Figure 7**

To formally test the normality of the forecast error, Shapiro-Wilk test is performed and the result is summarized in Table 5. The Shapiro-Wilk test statistics of 0.97643 obtained a *p*-value of 0.8221 which is larger than 0.05 level of significance. This implies that the null hypothesis cannot be rejected and that the forecast errors are not normally distributed. Therefore, the forecast errors behave like a Gaussian white noise.

**Table 5.** Shapiro-Wilk test for forecast errors

| Test statistic | $p$-value |
|---|---|
| 0.97643 | 0.8221 |

Using the $\mathrm{SARIMA}(0, 1, 1) \times (1, 0, 0)_{12}$ model, the forecasted values from January up to December 2019 are computed along with its 95% confidence intervals and the values are presented in Table 6. These forecasted values show that the least possible cases of HIV will occur in the month of July and the largest possible cases will occur in the month of October.

**Table 6.** Forecast values

| Month | Forecast value | Lower 95% CI | Upper 95% CI |
|---|---|---|---|
| January | 961.5688 | 830.3761 | 1102.568 |
| February | 900.7721 | 762.6867 | 1050.568 |
| March | 918.5014 | 768.5022 | 1082.131 |
| April | 922.5881 | 762.5092 | 1098.214 |
| May | 933.1516 | 762.9553 | 1120.815 |
| June | 950.4344 | 769.869 | 1150.389 |
| July | 895.7773 | 712.8983 | 1099.940 |
| August | 971.8274 | 772.8460 | 1194.042 |
| September | 934.7690 | 732.5110 | 1162.160 |
| October | 981.6211 | 766.6740 | 1223.641 |
| November | 931.1269 | 715.4399 | 1175.777 |
| December | 903.2617 | 684.6782 | 1152.707 |